\documentclass[conference,a4paper]{IEEEtran}
\usepackage{xcolor}
\usepackage{balance}
\usepackage{booktabs}
\usepackage{amsmath}
\usepackage{cite} 
\usepackage{float}
\usepackage{makecell}
\usepackage{array}

\ifCLASSINFOpdf
  \usepackage[pdftex]{graphicx}
  
\else
  \usepackage[dvips]{graphicx}
\fi
\ifCLASSOPTIONcompsoc
 \usepackage[caption=false,font=normalsize,labelfont=sf,textfont=sf]{subfig}
\else
 \usepackage[caption=false,font=footnotesize]{subfig}
\fi
\usepackage{tikz}

\begin{document}
%
\title{Confocal Ellipsoidal Reflectors with Phased Array Vivaldi Antenna Source for Imaging Systems}

\author{\IEEEauthorblockN{
Mohammad Hossein Koohi Ghamsari\IEEEauthorrefmark{1},   
Mahyar Mehri Pashaki\IEEEauthorrefmark{2},   
Mehdi Ahmadi-Boroujeni\IEEEauthorrefmark{3},    
}                                     
\IEEEauthorblockA{\IEEEauthorrefmark{1}
Department of Electrical Engineering,
Sharif University of Technology,
Tehran, Iran,
mohammad.koohi@sharif.edu}
\IEEEauthorblockA{\IEEEauthorrefmark{2}
Department of Electrical Engineering,
Sharif University of Technology,
Tehran, Iran,
mahyar.mehri.p@gmail.com}
\IEEEauthorblockA{\IEEEauthorrefmark{3}
Department of Electrical Engineering,
Sharif University of Technology,
Tehran, Iran,
ahmadi@sharif.edu}
}



\maketitle

\begin{abstract}
In this paper, an on-axis dual-reflector confocal ellipsoidal structure is presented for near-field imaging systems. In the proposed structure, the backscattered electromagnetic wave problem, known as the blockage effect, is reduced considerably using an elaborate design of the sub-reflector and precise alignment of the reflectors. The proposed geometry is analyzed, followed by a design example for the stand-off distance of 2 m. The blockage reduction characteristic is verified using ray-tracing simulation. Next, the scanning performance of the structure is investigated utilizing a Vivaldi phased array antenna as the source designed at the central frequency of 28 GHz. The full-wave simulations proved a field-of-view (FoV) of approximately 40 cm. Furthermore, tuning the proposed reflectors configuration stand-off distance is examined with a point source. The ray-tracing simulations showed that stand-off distance can be easily changed up to tens of centimeters with just a few centimeters of source point lateral displacement.
\end{abstract}

\vskip0.5\baselineskip

%

\section{Introduction}
Nowadays, with the increasing expansion of telecommunication and imaging applications, designing suitable antenna structures is a crucial requirement \cite{balanis2024evolution,shi2023broadband,jiang2024millimeter,balanis2016antenna}. The advantages of reflector antenna structures and reflective surfaces such as high gain and efficiency, wide-band operation, and a low sidelobe level (SLL), make these structures highly desirable for numerous applications in a wide range of electromagnetic spectrum, from microwave frequencies to millimeter-wave (mm-wave) and terahertz (THz) \cite{baghel2024novel,martinez2024multibeam,vuyyuru2023efficient,ghamsari2024design,sano2024reflection,ghamsari2022confocal,sanchez2024double}. 

Reflector antenna systems have been designed for a wide range of applications. Reflector configurations are extensively utilized in telecommunications, facilitating long-distance signal transmission and reception with minimal interference \cite{rao2013handbook,phan2024sub,taillieu2024low}. In radio astronomy, reflector antennas enable the collection of faint signals from distant celestial bodies, enhancing our understanding of the universe \cite{baars2007paraboloidal,soliman2016optimization}.  Additionally, reflector systems play a crucial role in radar technology, contributing to applications such as airport security and surveillance by providing precise target detection and tracking capabilities \cite{toshev2010analysis,diao2024poynting}. Furthermore, reflectors are increasingly employed in satellite communication, where their ability to form multiple isolated beams allows for efficient frequency reuse and improved coverage \cite{robustillo2024spherical}.

In recent years, reflector antenna systems have played a crucial role in mm-wave and THz imaging systems by providing high gain, improved spatial resolution, the ability to effectively control beam patterns, and the ability to scan the focal plane easily, making them essential for high-resolution imaging systems \cite{rappaport2024wideband,sheen2009active,wang2024phase}. Specifically, Gregorian-based reflector systems have attracted great attention for imaging in the near-field (Fresnel region), for scenarios such as focal plane imaging, due to their ability to control aberrations and thus provide acceptable scanning capabilities \cite{ghamsari2022design}. For example, a confocal Gregorian reflector system is designed for rapid scanning and refocusing of a THz beam for high-resolution stand-off imaging \cite{llombart2010confocal}. The proposed system achieves effective beam scanning over 0.5 m at a 25 m stand-off range, validated through numerical simulations while maintaining a minimal increase in beamwidth. The zooming and scanning capabilities of a Gregorian confocal dual reflector antenna, utilizing a planar feed array and active mechanical deformation to compensate for quadratic aberrations discussed \cite{martinez2008zooming}. A method to enhance THz imaging resolution in a dual reflector system is presented in \cite{zhou2018bifocal}, verified with a Gregorian system at 220 GHz, achieving better than 3 cm resolution over a 50 cm by 100 cm field-of-view (FoV) at 8 m, enabling active THz imaging of the human body.

However, there are still some drawbacks and challenges both in the design and performance of these reflector structures. First, many reflector configurations require large dimensions and complex designs, especially with multiple reflectors, which can complicate manufacturing and alignment processes. Second, quadratic aberrations may occur, necessitating active mechanical deformation to correct them, which adds to the complexity of the system. Third, reflector antennas can suffer from cross-polarization effects that degrade overall performance. Fourth, blockage can lead to reduced radiation efficiency, decreased signal-to-noise ratio (S/N), and diminished imaging systems detector sensitivity due to interference from return signals. Furthermore, the maximum scanning range can be restricted by blockage effects, impacting performance in many applications.

Various design techniques are proposed in the literature for improving the reflector antenna structures. Shaping and deforming reflectors using iterative approaches and asymmetric or off-axis techniques have been widely used for various optimization goals \cite{eichenberger2018deformable,kosulnikov2024experimental}. However, each of these approaches has its own disadvantages. For instance, utilizing asymmetric or off-axis configurations for reducing the blockage increases the overall system size and complexity in reflector manufacturing and alignment, as well as increasing the cross-polarization effects.

In this paper, a simple on-axis dual-reflector confocal ellipsoidal configuration with a Vivaldi phased array antenna as the feeding source is proposed for imaging in the Fresnel region. This structure is obtained by applying an elaborate modification to the traditional dual-reflector Gregorian structure. With these changes, the overall blockage effect is decreased considerably. Moreover, the scanning performance and tunning of the stand-off distance have been verified using full-wave and ray tracing simulations.

\section{CONFOCAL ELLIPSOIDAL STRUCTURE DESIGN}
In this section, first, the general geometry of the proposed confocal ellipsoidal reflector system is presented and analyzed. Next, a design example is presented to verify the blockage reduction characteristic of our design by comparing it with a traditional Gregorian antenna structure. 

\subsection{General Geometry}
The general structure of the proposed symmetric dual-reflector configuration is depicted in Fig.~1. Similar to the standard Gregorian structure, a main reflector and sub-reflector are utilized with dimensions $D_M$ and $D_S$, respectively. In comparison to reflector systems for telecommunication applications where the main reflector is a parabolic curve, here, the main reflector is a standard elliptical conic curve that is decentered vertically by $D_B/2$, leaving a hole in the middle of the geometry. $P_2$ indicates the second focal point of the main reflector, located at the distance $d_s$ in the focal plane. Also, $F$ denotes the focal length of the main reflector.

The sub-reflector is also an elliptical conic curve that is simultaneously displaced and tilted counter-clockwise to share a common focus with the main reflector (point $P_1$) and the source (point $O$). The sub-reflector surface is produced from an ellipse with eccentricity $e$ and interfocal distance $2c$. Symbols $\theta_U$ and $\theta_L$ show the upper and lower angles of the main reflector, respectively. Notation $\theta_E$ represents the sub-reflector edge angle and $\beta$ is the tilt angle between the symmetry axis and ellipse axis. Also, $V_M$ and $V_S$ denote the y-coordinates of the points on the main reflector and sub-reflector, respectively, corresponding to the principal ray of the geometry.

Based on Fig. 1, the main reflector and the sub-reflector curves can be generally described as

\begin{equation}
	\overline{P_1 M} = \frac{2F}{1 + \cos\theta_m} \label{eq1}\\
\end{equation}

\begin{equation}
	\overline{OS} \mp \overline{SP_1} = \frac{2c}{e} \label{eq2}\\
\end{equation}

where $\overline{P_1 M}$ is the distance between the point $P_1$ and point $M$, $\overline{OS}$ is the distance between the focal point $O$ and the point $S$ on the sub-reflector surface. Similarly, $\overline{SP_1}$ is the distance between point $S$ and point $P_1$. Furthermore, from the basic trigonometric laws, it can be easily observed that
\begin{equation}
	\tan{\theta_U} = \frac{-D_M}{2(V_S - V_M)} \label{eq3}\\
\end{equation}

\begin{figure}[!t]
	\centering
	\includegraphics[scale=0.55,trim=2.5cm 18cm 2.5cm 3.5cm,clip=true]{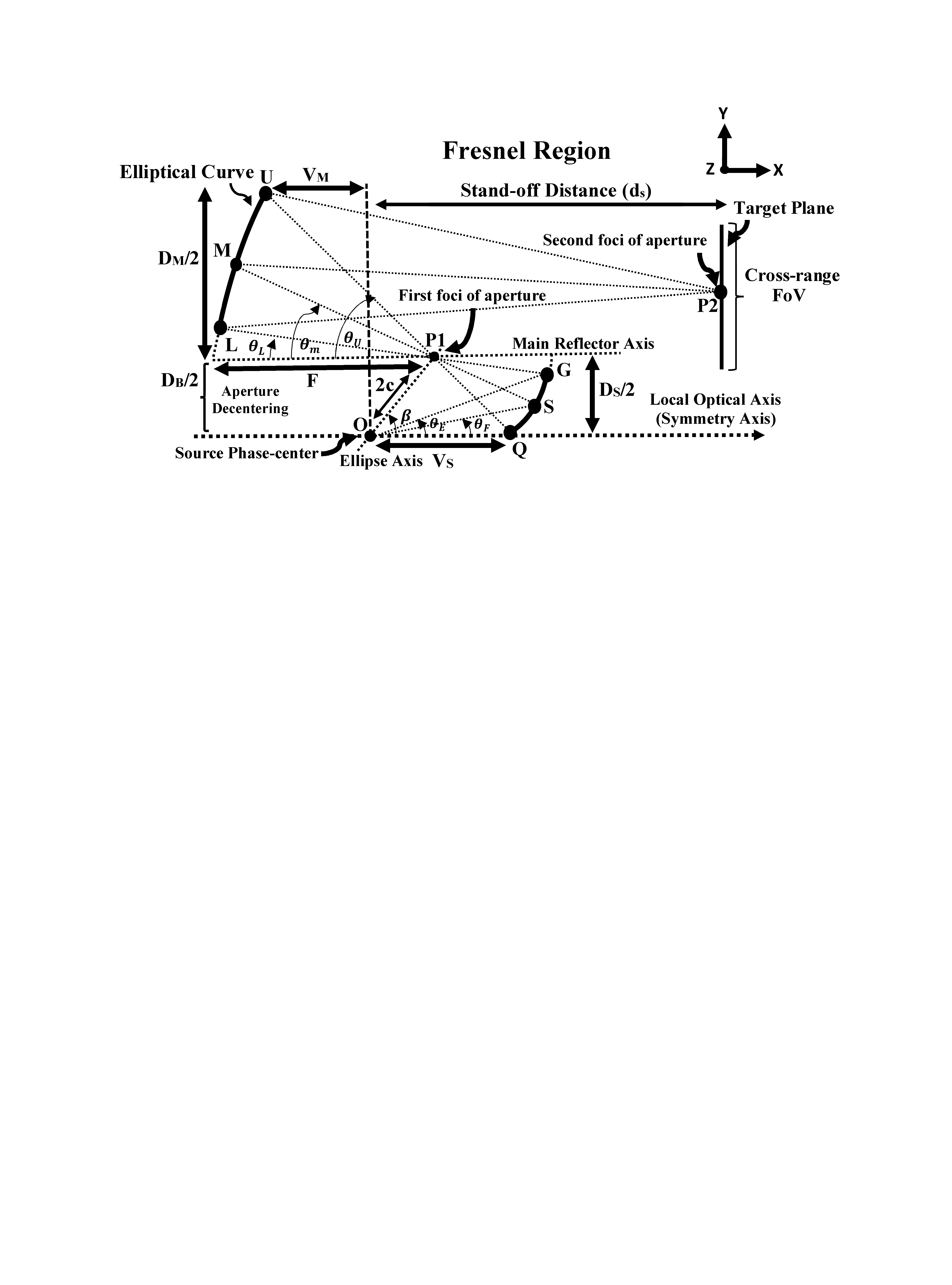}
	
	\captionsetup{
		font=footnotesize,        
		singlelinecheck=false,    
		margin={0pt,0pt},         
		skip=0pt,                 
		belowskip=20pt            
	}
	
	\caption{Symmetric dual-reflector confocal ellipsoidal configuration. The modified displaced sub-reflector reflects the emanated rays from the source toward the main reflector. The elliptical main reflector, aperture, converges the divergent rays coming from the sub-reflector to the point P2, the second foci of the aperture at the stand-off distance.}
	\label{fig1}
\end{figure}

The phase center of the electromagnetic source is located at point O. The basic geometrical optics and ray-tracing principles can be used to analyze the structure. First, the emanated divergent rays of the source incident on the exterior surface of the modified sub-reflector. Then, after reflection, the rays converge to point P1, the first foci of the main reflector elliptical curve, and again diverge to the interior surface of the main reflector. Finally, the elliptical main reflector focuses the rays reflected from the sub-reflector at the Fresnel region at point P2, the second focal point of the main reflector. Moreover, the field-of-view (FoV) can be defined as the maximum range of scanning the focus point within an acceptable range of aberrations and before the rays become increasingly divergent.

\subsection{Design Example}
A design example is presented in this section to illustrate the ray's paths and blockage behavior of the structure. The design specifications are reported in Table I. According to this Table, the stand-off distance, where rays are focused, is considered 200 cm, and a FoV of approximately 40 cm is expected. Also, a standard Gregorian structure with the same specification is designed to compare the results.

\begin{table}[b]
	\centering
	\caption{Design Parameters of the Confocal Ellipsoidal Structure}
	\begin{tabular}{|c|c|c|c|c|}
		\hline
		\textbf{$D_M$} & \textbf{$D_S$} & \textbf{$D_B$} & \textbf{$d_s$} & \textbf{$FoV$} \\ 
		\hline
		80 cm    & 35 cm           & 16 cm           & 200 cm            & 40 cm            \\ 
 
		\hline
	\end{tabular}
	\label{tab1}
\end{table}

The ray-tracing module of the COMSOL Multiphysics is adopted to perform the simulations. The obtained results are shown in Fig. 2. In this figure, the rays are colored based on the optical path length (OPL). It can be noticed that while no rays are blocked in the structure of Fig. 2 (a), a significant number of rays are blocked in the Gregorian structure of Fig. 2 (b), which proves the improved blockage performance of the proposed structure.
\begin{figure}[!t]
	\centering
	
	\includegraphics[scale=0.7,trim=4.5cm 7cm 2.5cm 3.5cm,clip=true]{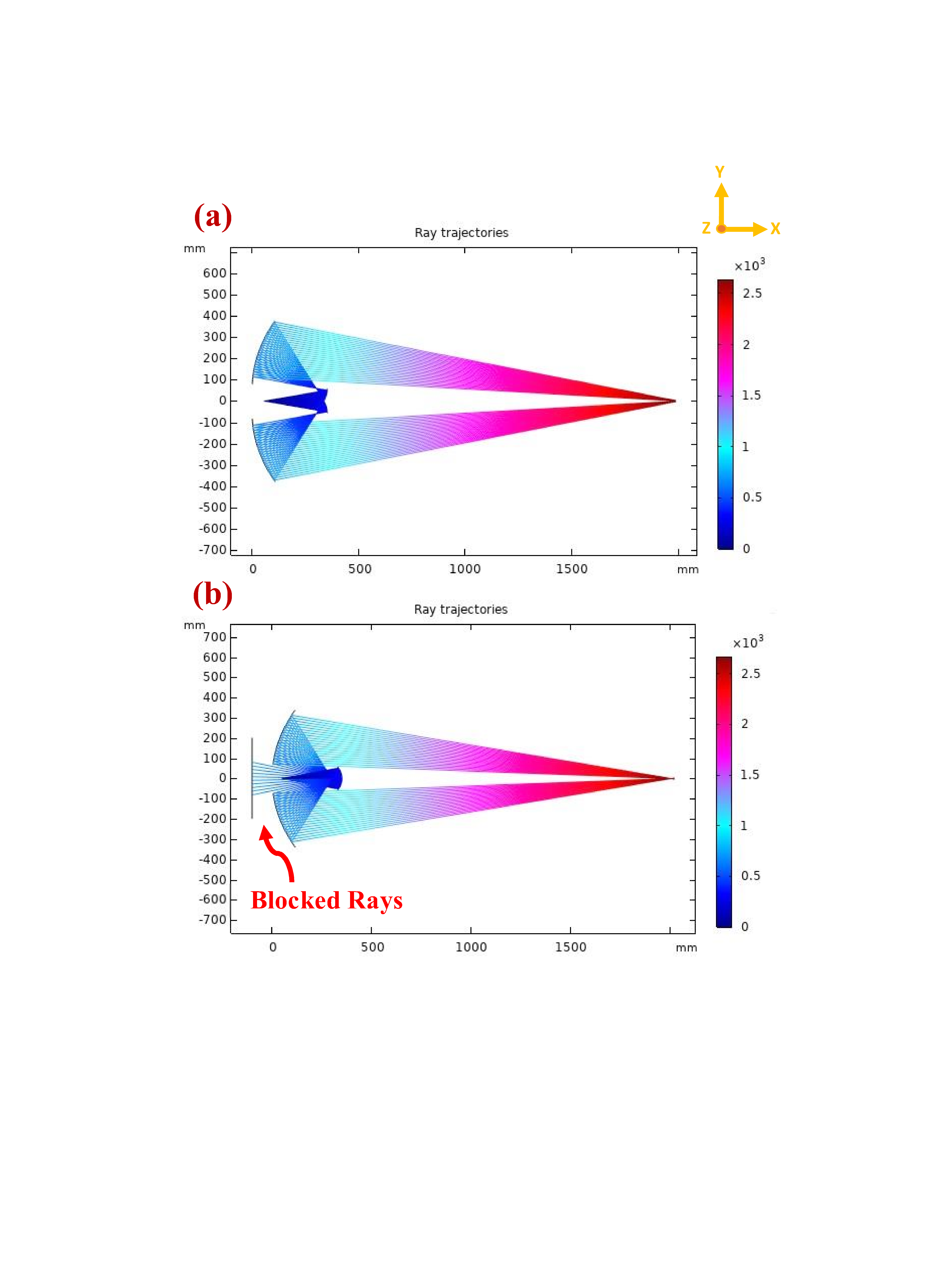}
	
	\captionsetup{
		font=footnotesize, 
		singlelinecheck=false, 
		margin={0pt,0pt} 
	}	
	\caption{Comparing the ray-tracing simulation results of the proposed structure and standard Gregorian. (a) Simulated rays’ performance of the proposed dual-reflector confocal ellipsoidal structure. (b) Simulated rays’ performance of the standard Gregorian structure. The color bar in both insets shows the optical path length (OPL) of the rays and the structure source point (O) is considered as the reference for the OPL calculations.}
	\label{fig2}
\end{figure}

\section{EVALUATING IMAGING PERFORMANCE}
The scanning performance and the tuning of the stand-off focusing distance, which are two critical abilities of an optical setup for imaging applications, are investigated in this section.

\subsection{Scanning Performance}
In this section, the capability of the proposed structure to scan in the near-field is examined. First, it is useful to have an intuition of the electric field (E-field) intensity distribution in the Fresnel region and also verify the focusing ability of the structure in a full-wave simulation. Therefore, the Electromagnetic Waves module of the COMSOL Multiphysics is utilized. As shown in Fig. 3, a point source is defined at point O of Fig. 1 at the frequency of 30 GHz. On the right side of this figure, the E-field intensity is maximized in a linear region, or E-field caustic. This is where the object can be located or displaced in an imaging scenario. The FoV can be defined as the maximum range of scanning the focus point within an acceptable value for the half-power-beamwidth (HPBW).
\begin{figure}[!t]
	\centering
	
	\includegraphics[scale=0.4,trim=5cm 3cm 5cm 3cm,clip=true]{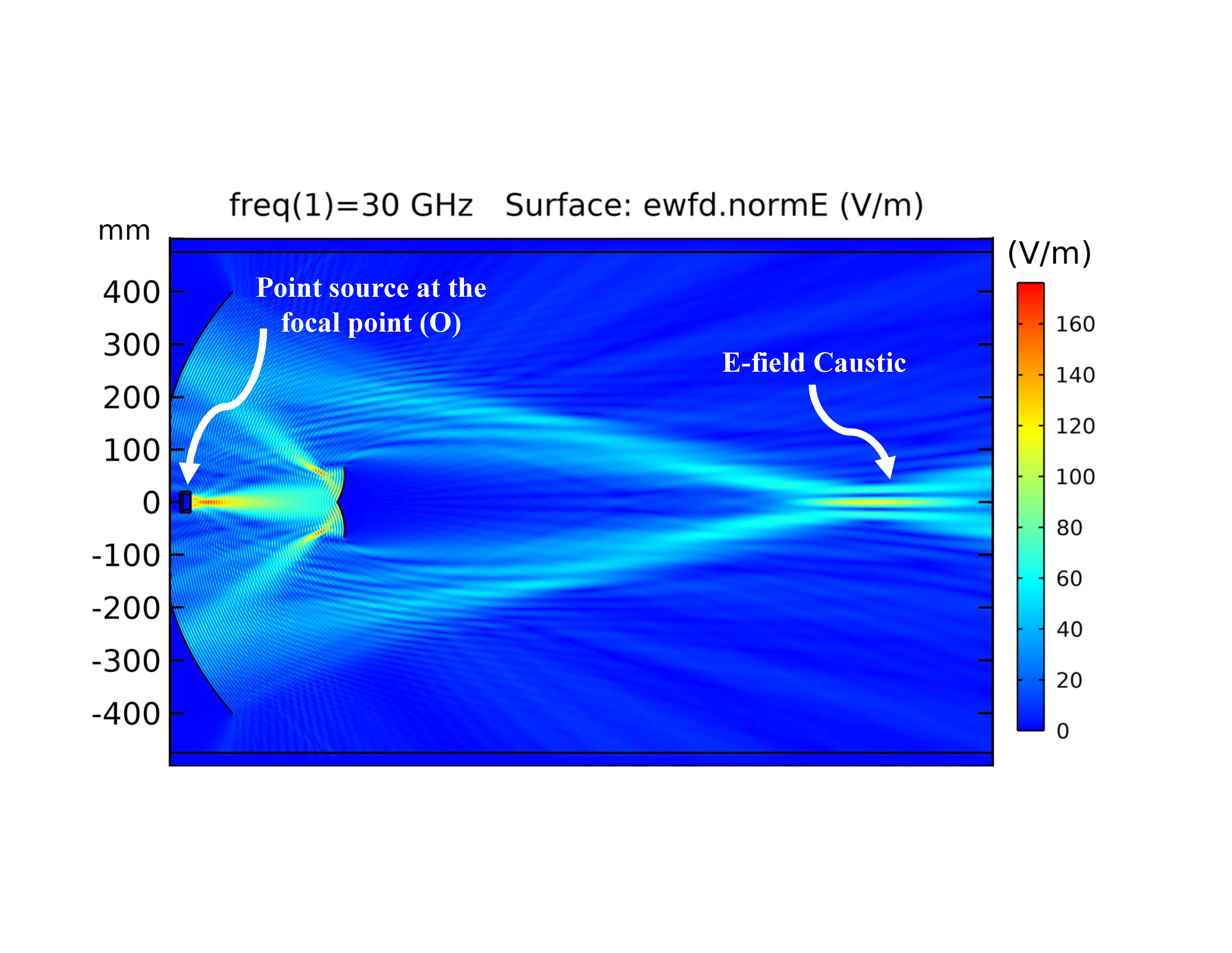}
	
	\captionsetup{
		font=footnotesize, 
		singlelinecheck=false, 
		margin={0pt,0pt} 
	}	
	\caption{Full-wave simulation of the dual-reflector confocal ellipsoidal Fresnel region.}
	\label{fig3}
\end{figure}
To investigate the scanning performance of the proposed reflector configuration and calculate the FoV in a more practical approach, a phased array Vivaldi antenna is designed as the source of the structure for the central frequency of 28 GHz and beam steering capability from -30 to +30 degrees. The unit cell of this antenna is depicted in Fig.~4 (a) and the geometrical specifications are listed in Table~II. The reflection coefficient of the unit cell is simulated using CST Studio Suite and is illustrated in Fig.~4 (b), displaying a 10-dB bandwidth ranging from 26 to 30 GHz. The overall structure of the antenna consisting of a 4×4 array of the proposed unit cells is also shown in Fig.~4 (c).

	\begin{table}[b]
		\centering
		\caption{Geometrical parameters of the Vivaldi antenna unit cell.}
		\begin{tabular}{|c|c|c|c|} 
			\hline
			\textbf{Parameter} & \textbf{Value (mm)} & \textbf{Parameter} & \textbf{Value (mm)} \\ \hline
			d & 4.46  & $L_{s}$ & 7.6 \\ \hline
			h & 2 & $L_{1}$ & 2.66 \\ \hline
			t & 2.28 & $L_{2}$ & 1 \\ \hline
			w & 5.7 & $L_{3}$ & 2 \\ \hline
			$w_{a}$ & 4.75 & $L_{4}$ & 1.15 \\ \hline
			$w_{t}$ & 0.45 & $L_{5}$ & 2 \\ \hline
			L & 14.25 & $L_{6}$ & 2 \\ \hline
		\end{tabular}
		\label{tab2}
	\end{table}

\begin{figure}[t]
	\centering
	\begin{tikzpicture}
		
		\node[anchor=south west] at (0,0) {\includegraphics[scale=0.4,trim=8.5cm 3cm 14cm 1cm,clip=true]{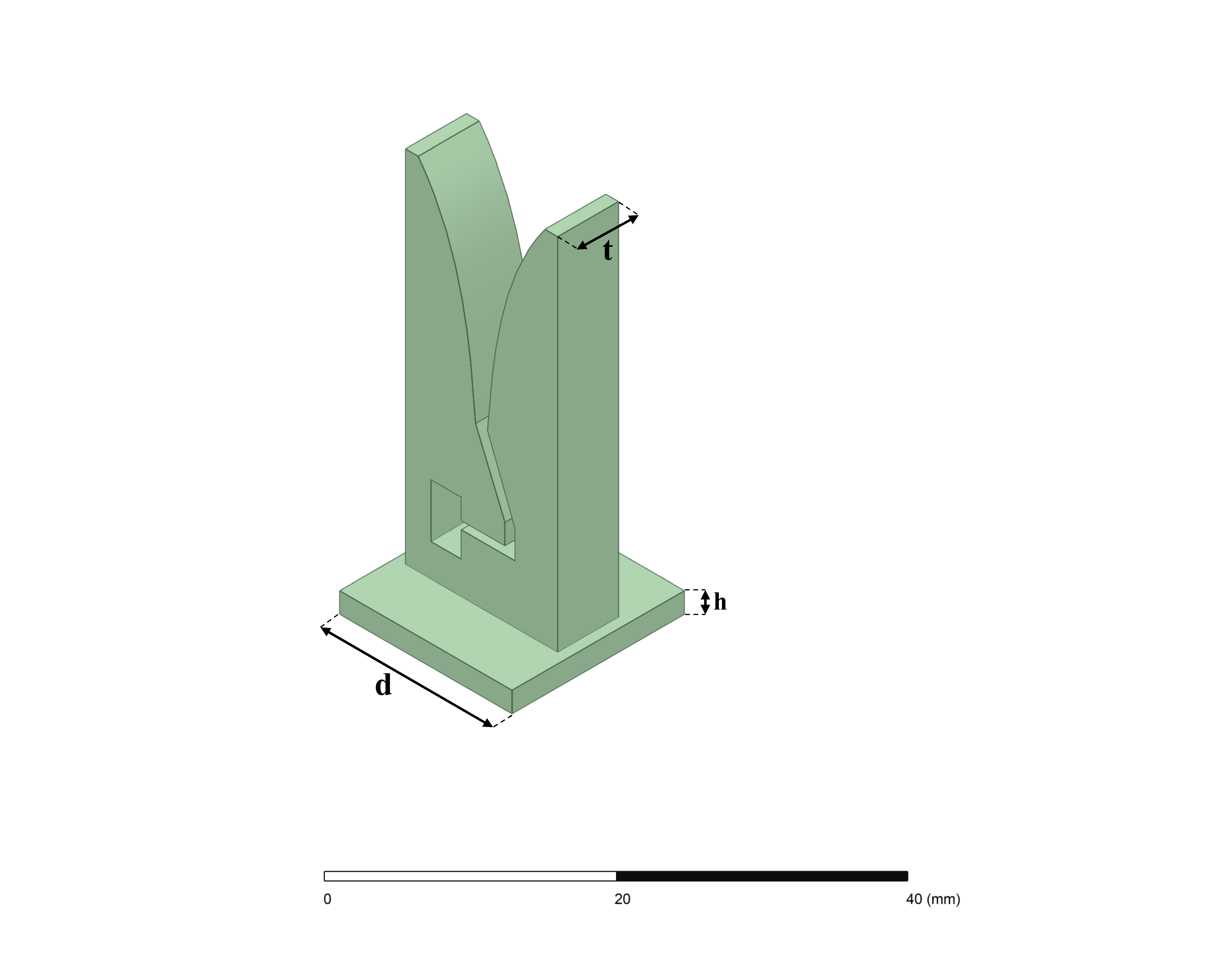}};
		\node[anchor=south west] at (5,0) {\includegraphics[scale=0.34,trim=11cm 1.5cm 5cm 0cm,clip=true]{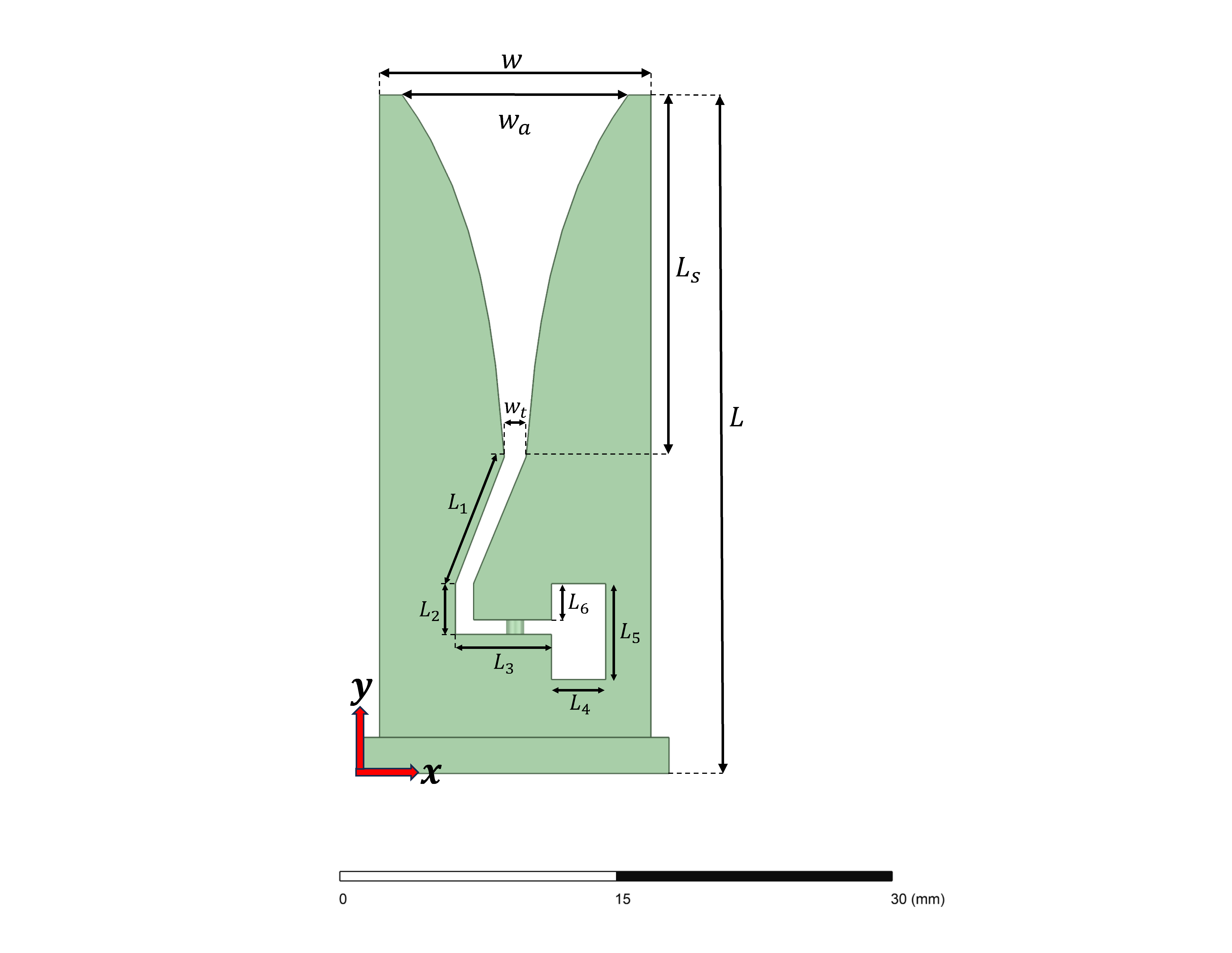}};
		
		\node at (1,3.5) {\textbf{\large (a)}};  
		\node at (6.5,3.5) {\small };  
		
		\node[anchor=south west] at (0,-5) {\includegraphics[scale=0.32,trim=3.8cm 8cm 4.5cm 8cm,clip=true]{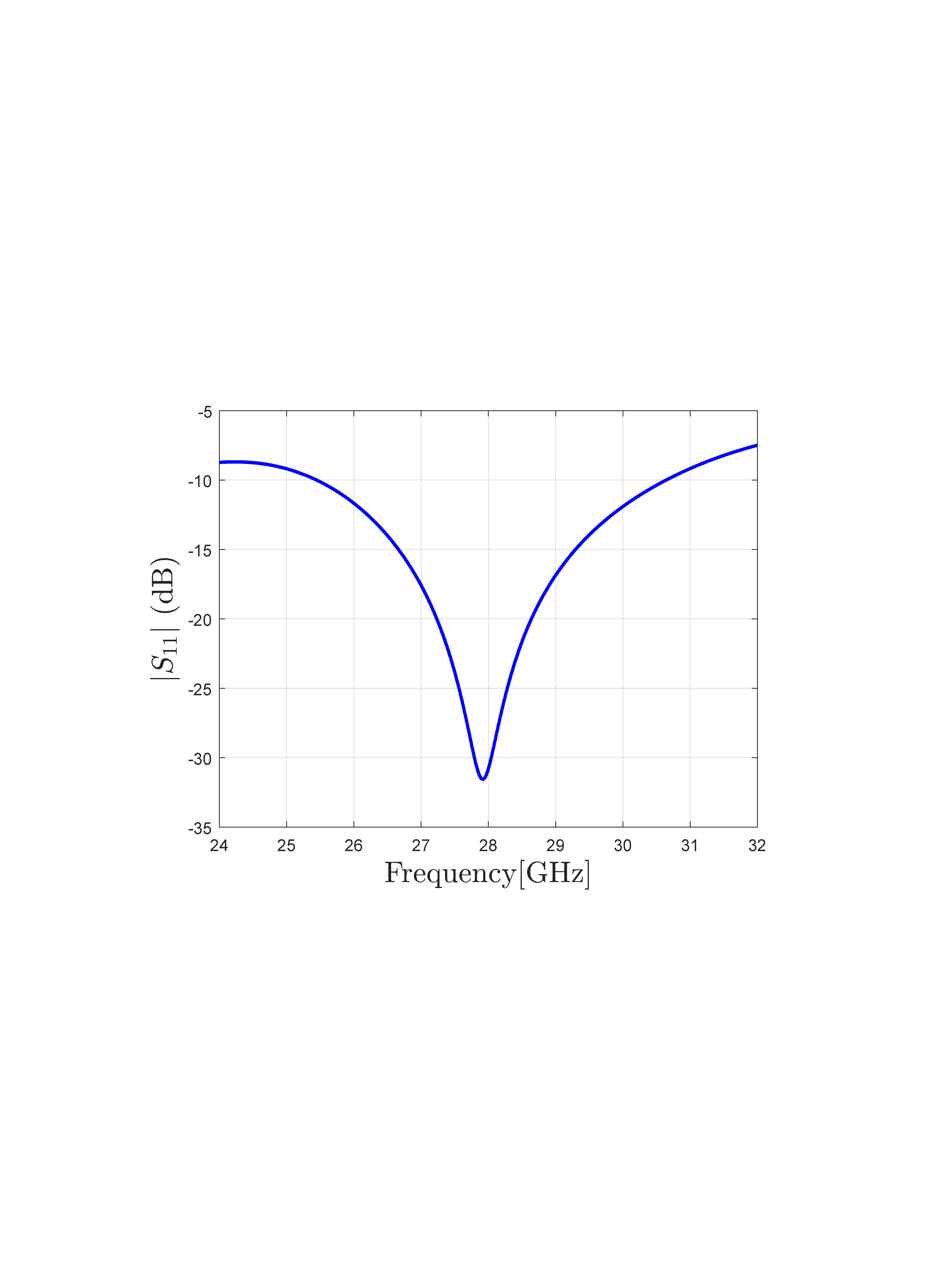}};
		\node[anchor=south west] at (5,-5) {\includegraphics[scale=0.027,trim=62cm 20cm 60cm 9.5cm,clip=true]{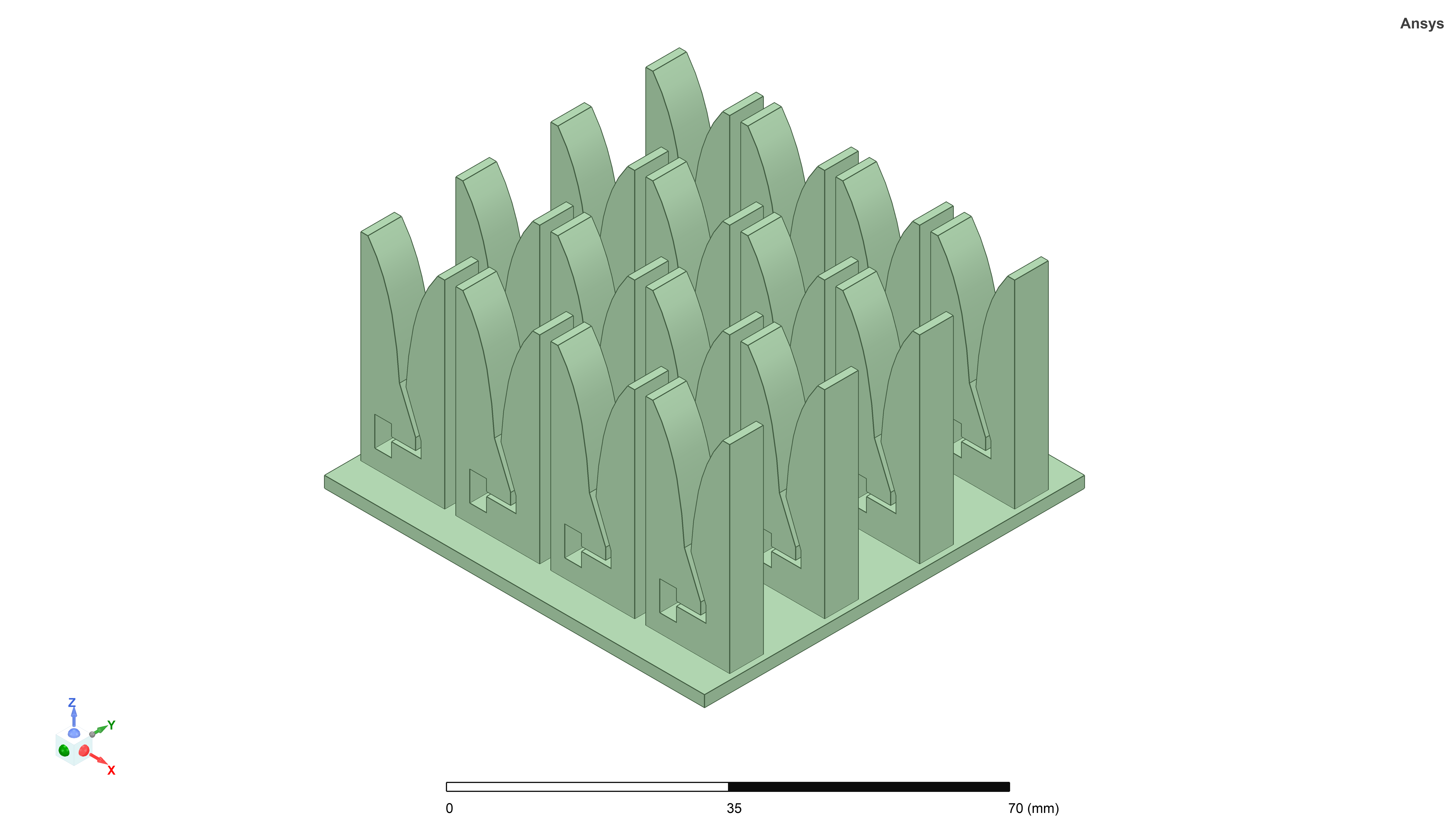}};
		
		\node at (2.4,-1) {\textbf{\large (b)}};  
		\node at (6.8,-1) {\textbf{\large (c)}};  
		
	\end{tikzpicture}
	
	\captionsetup{
		font=footnotesize, 
		singlelinecheck=false, 
		margin={0pt,0pt} 
	}
	\caption{Phased array Vivaldi antenna. (a) The proposed unit cell of the Vivaldi antenna. (b) Simulated reflection coefficient of the Vivaldi antenna unit cell. (c) Overall antenna array structure is composed of a 4×4 array of unit cells.}
	\label{fig4}
\end{figure}
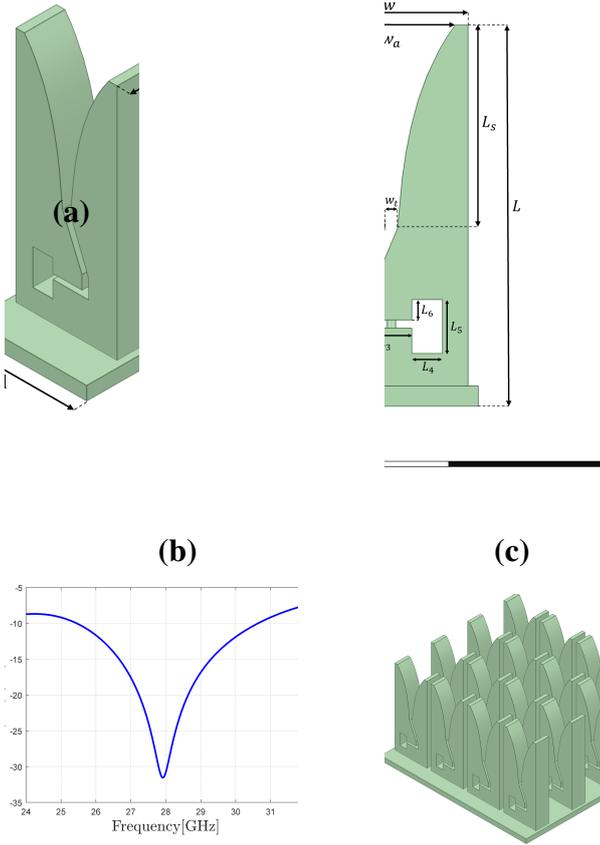
	
Using this Vivaldi phased array antenna in the dual-reflector structure as the source, the scanning performance of the structure is obtained as shown in Table III. Based on this table, by changing the phased array antenna beam angle by up to 30 degrees, the stand-off focusing point is scanned to about 19.7 cm, giving an approximate 40 cm desired FoV of Table I.

\begin{table}[b]
	\setlength{\tabcolsep}{0pt}
	\centering
	\caption{Scanning performance of the reflector structure.}
	\begin{tabular}{ |>{\centering\arraybackslash}m{2.3cm}|>{\centering\arraybackslash}m{0.7cm}|>{\centering\arraybackslash}m{0.7cm}|>{\centering\arraybackslash}m{0.8cm}|>{\centering\arraybackslash}m{0.8cm}|>{\centering\arraybackslash}m{0.8cm}|>{\centering\arraybackslash}m{0.8cm}| } 
		\hline
		\textbf{\makecell{Phased array \\ scanning (°)}} & $\mp$5 & $\mp$10 & $\mp$15 & $\mp$20 & $\mp$25 & $\mp$30 \\ \hline
		\textbf{\makecell{Stand-off distance \\ scanning (cm)}} & $\mp$3.5 & $\mp$7.6 & $\mp$10.3 & $\mp$14.9 & $\mp$18.6 & $\mp$19.7 \\ \hline
	\end{tabular}
	\label{tab3}
\end{table}

\subsection{Tunning Stand-off Distance}
The focusing distance of the rays is an important specification for any imaging system. However, it is necessary for many applications to change this distance based on the imaging setup. Therefore, the dynamic tunning ability of the focusing distance, technically known as refocusing, is considered a critical advantage. In the proposed reflector system, the stand-off focusing distance is the second focal point of the main elliptical reflector. The equations governing the focal points of the main reflector are
\begin{equation}
\begin{aligned}
	d_1 &= \frac{r}{p} \left( 1 - \sqrt{1 - p} \right) \\
	d_2 &= \frac{r}{p} \left( 1 + \sqrt{1 - p} \right)
\end{aligned}
\end{equation}

where \(d_1\) and \(d_2\) are the first and second focal lengths of the main reflector, respectively, and the relation \(d_2 = d_1 + 2c\) is always satisfied. Also, parameters \(r\) and \(p\) in these relations are the radius and the conic constant of the main reflector elliptical curve, respectively. Based on these relations and Fig. 5, the stand-off distance of the proposed structure can be easily tuned by the lateral displacement of the source point. To be more specific, by changing the feeding point about \(\Delta_{fx}\), the stand-off focusing point will also change about \(\Delta_{sx}\). Some examples of these changes are listed in Table IV. In this table, \(M_S\) is the reflector system magnification factor and is defined as the ratio of the main reflector's first focal length to the sub-reflector's first focal length.

\begin{figure}[t]
	\centering
	
	\includegraphics[scale=0.55,trim=7.2cm 10cm 6.8cm 3cm,clip=true]{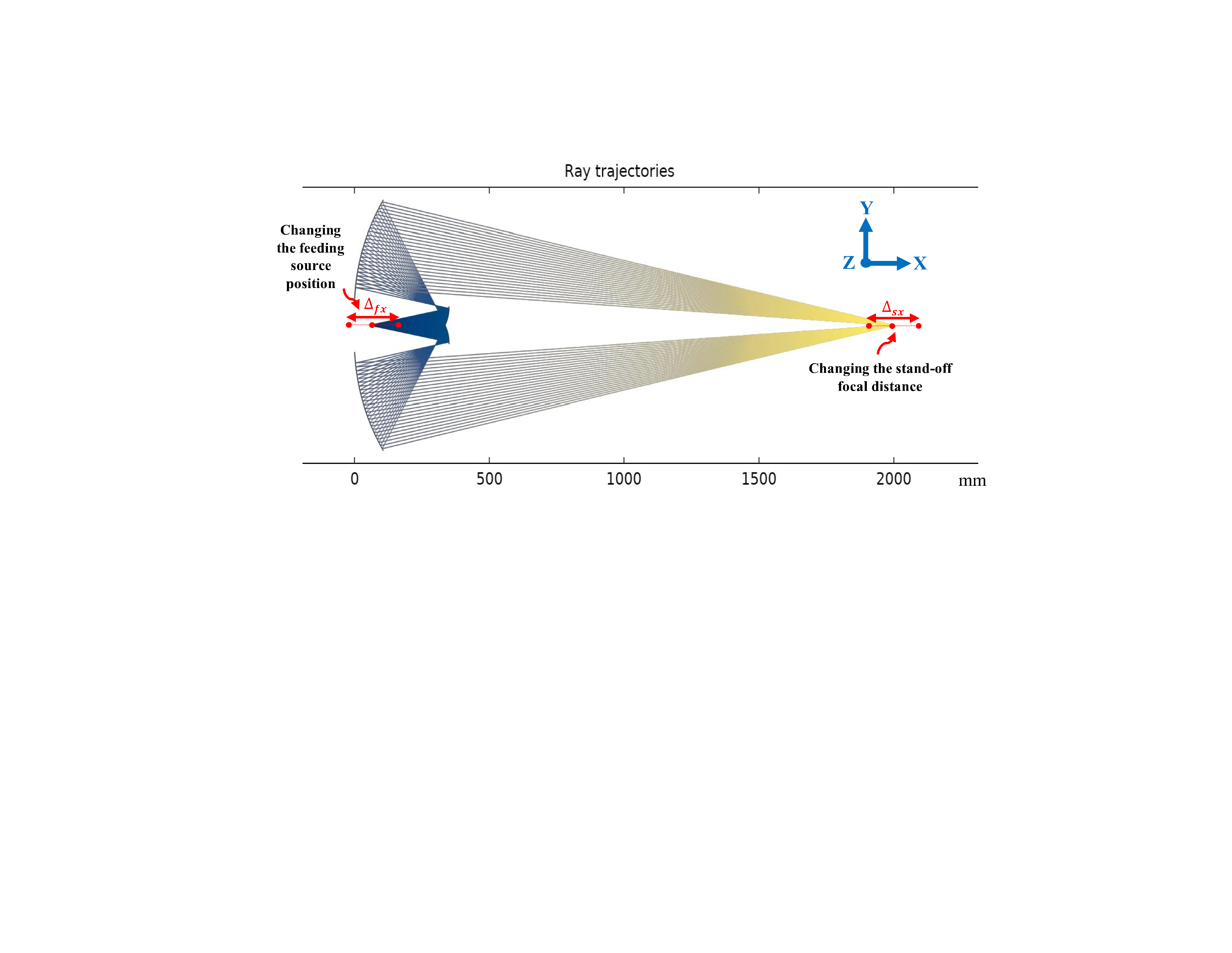}
	
	\captionsetup{
		font=footnotesize, 
		singlelinecheck=false, 
		margin={0pt,0pt} 
	}	
	\caption{Refocusing performance of the proposed structure by displacing the feeding source position laterally.}
	\label{fig5}
\end{figure}

\begin{table}[t]
	\setlength{\tabcolsep}{7pt}
	\centering
	\caption{Stand-off distance tuning using feed displacement and magnification factor.}
	\begin{tabular}{|c|c|c|c|c|c|c|} 
		\hline
		\textbf{$\Delta_{fx}$ [cm]} & -2.5 & -1.5 & -0.5 & 3.5 & 1.5 & 0.5 \\ \hline
		\textbf{$\Delta_{sx}$ [cm]  ($M_{s}=6.5$)} & 62 & 28 & 10 & -48 & -27 & -13 \\ \hline
		\textbf{$\Delta_{sx}$ [cm]  ($M_{s}=2.5$)} & 108 & 66 & 27 & -99 & -69 & -25 \\ \hline
		\textbf{$\Delta_{sx}$ [cm]  ($M_{s}=11$)} & 33 & 22 & 12 & -26 & -12 & -6 \\ \hline
	\end{tabular}
	\label{tab:my_label}
\end{table}

Based on Table IV, the stand-off distance has an inverse relation with the magnification factor. For example, by decreasing the magnification factor from 6.5 to 2.5, $\Delta_{sx}$ is increased from 62 cm to 108 cm for $\Delta_{fx} = -2.5$ cm. Furthermore, this table shows that by displacing the source point just a few centimeters, the stand-off focusing point can be tuned a few tens of centimeters. Additionally, these values are strongly dependent on the magnification factor of the reflector structure.
\balance

\section{CONCLUSION}
In this paper, a confocal ellipsoidal reflector system for near-field imaging applications is proposed. By effectively reducing the blockage effect through meticulous design and alignment of the reflectors, this configuration enhances the overall scanning performance and flexibility in tuning the focusing distance. Utilizing a Vivaldi phased array antenna and performing full-wave simulations, the capability of the presented structure to achieve a FoV of approximately 40 cm is demonstrated. The ray tracing simulation results presented dynamic adjustments in stand-off distance with minimal lateral displacement of the source point. Moreover, considering the reflector antenna frequency independence, the proposed structure can be effectively used in a wide range of the electromagnetic spectrum, from microwave and mm-wave imaging systems to THz imaging systems, with easily tunable properties for each imaging scenario.

\bibliographystyle{IEEEtran}
\bibliography{./mybib}

\end{document}